Pasquale Lisena, Manel Achichi, Pierre Choffé, Cécile Cecconi, Konstantin Todorov, and Bernard Jacquemin and Raphaël Troncy

# Improving (Re-) Usability of Musical Datasets: An Overview of the DOREMUS Project

**Abstract:** DOREMUS works on a better description of music by building new tools to link and explore the data of three French institutions. This paper gives an overview of the data model based on FRBRoo, explains the conversion and linking processes using linked data technologies and presents the prototypes created to consume the data according to the web users' needs.

**Verbesserung der (Wieder)Verwendbarkeit von Musikdaten: Ein Überblick über das DOREMUS Projekt**

**Zusammenfassung**: Das DOREMUS Projekt strebt eine bessere Beschreibung von Musik an, indem es Daten dreier französicher Institutionen untersucht und zusammenführt. Der vorliegende Artikel gibt einen Überblick über das auf FRBRoo basierende Datenmodell, das die automatische Umwandlung und Verlinkung von Daten ermöglicht. Er stellt Prototypen vor, wie die Daten nach den Bedürfnissen der Webnutzer verarbeitet werden können.

**Schlüsselwörter:** FRBRoo, Musikwerke, semantisches Web, Datenumwandlung, Data Linking, Musikpraktiken, Empfehlungssystem

## 1    Introduction

"I would like to listen to the original version of *Night on Bald Mountain* by M. Mussorgsky, not the usual orchestral one by Rimski-Korsakov. Do you have a record of it or do you know where I could attend a performance of it?"

To find a quick answer to this question on the web is not that easy, even for a musical librarian. Indeed, musical works and performances are complex information objects, especially in classical or traditional repertoire, due to their multilingual titles, various versions or interpretations and complex structures (mouvements, acts, etc.).

The linked data technologies offer opportunities to improve the description of music on the web by making it easier to link and reuse different sources of data. Several research projects are currently testing their potential, each of them exploring a different part of the musical world. The Listening Experience Database (LED) project[1] focuses on collecting information about people's experience of listening to music. The data comes from sources like the British National Bibliography, MusicBrainz or DBpedia but also from crowdsourcing. Fusing Audio and Semantic Technologies for Intelligent Music Production and Consumption (FAST)[2] aims to improve the workflow of music, from the production to the end listener. They work on metadata but also on audio signal processing techniques. The Linked Data for Production (LD4P) project[3] deals with library metadata and is developing domain-specific extensions to the BIBFRAME model – an ontology developed by the Library of Congress from the FRBR model – like Performed Music[4] or Moving Images.[5]

The research undertaken by the DOREMUS project since 2014 is complementary to the ones cited above. It aims to link the data of three French musical institutions and to provide new ways of exploring them, according to the needs of the users. The work is focused on making the best of existing metadata, not dealing with audio signal analysis or crowdsourcing. The project brings together experts from various professional backgrounds: libraries (Bibliothèque nationale de France (BnF), Radio France (RF) and Philharmonie de Paris (PP)), computer science laboratory (LIRMM), information and communication technologies laboratory (Eurecom), information science and communication laboratory (GERiiCO) and two private companies (Ourouk, Meaning engines).

In this paper, we give an overview of the project with emphasis on the FRBR-oriented data model, the linking and aligning processes, and the prototypes created to consume data. In Section 1, we present the original datasets, the data model of the project – the DOREMUS extension of FRBRoo – and how the datasets are converted to the RDF format. In Section 2, we present how they are linked and how we will try to refine their structure and to link them with other datasets. In Section 3, we present how, thanks to an analysis of users' needs, we work on new ways to visualize and use these data.

## 2    Works and Events at the Heart of the Project

### 2.1    Original Datasets, Authorities and Controlled Vocabularies

The data used in the project contains descriptions of different entities related to music: documents (books, music scores, recordings...) as well as immaterial objects (musical works and performances).

---

These descriptions are heterogeneous since they are coming from six databases held by institutions with various roles and audiences: BnF, Radio France and the Philharmonie de Paris. An overview is provided in Table 1.

**Table 1:** Content of the original datasets from the BnF, Philharmonie de Paris and Radio France

| | BnF | Philharmonie | | Radio France | | |
|---|---|---|---|---|---|---|
| | Catalogue général | Médiathèque | Events | Discothèque | Documentation musicale | Documentation sonore |
| | XML / INTERMARC | XML / UNIMARC or INTERMARC | XML | XML | XML | XML |
| Uniform titles and musical works entries | 135 940 | 156 159 | | | 62 550 | |
| Scores | 89 184 | 30 319 | | | 9 154 | |
| Books | | 21 035 | | | | |
| Recordings (audio & visual) | 156 159 | 11 049 | | 340 609 | | |
| Performances | | | 2 717 | | 7 700 | 1 800 |

According to their origin, the datasets may differ in many ways such as granularity, format, or structure. By converting them into a single and consistent format while preserving their expressiveness is one of the principal issues of the DOREMUS project and a prerequisite for any interlinking attempt.

Though the structure of the databases is very different, each one of them uses a system of authorities to link its records. These in-house or standard lists of persons, corporate bodies, or musical concepts will play a vital role in the interlinking process, facilitating the building of bridges between the datasets. Most of them are multilingual and cover subjects such as musical keys, instruments and voices, modes, musical genres, ethnic groups, geographical places, historical periods, topical descriptors and functions. New lists will be created during the project like one about thematic catalogues of classical compositions.

## 2.2 Building a Common FRBR-oriented Data Model
In order to achieve its goals, the DOREMUS project needed a data model based on the familiar concepts of the FRBRer model allowing the description of musical works and their publications but also the description of concerts, festivals, and recordings that are part of the activities of Radio France and the Philharmonie de Paris. The most widely used ontology for music (Music Ontology[6]) could not fulfill this precise requirement since our records cover basically the whole range of music, from popular songs to the most complex music, including the ones that the familiar FRBRer model is not good at describing. Hence, there is the necessity to build our own model.

The DOREMUS model was developed as an extension of the FRBRoo model, itself an extension of CIDOC CRM[7] and it can not be used alone. CIDOC-CRM – created for describing cultural heritage – is a dynamic model in which the notion of event informs the whole modeling process. The FRBRoo extension includes the four levels of the FRBRer description (Work, Expression, Manifestation and Item - also referred to as WEMI), while offering richer possibilities of description and an easier implementation. The DOREMUS model adds new elements to FRBRoo and CIDOC-CRM whenever needed to precisely express any music-related concept or relationship.

The three models follow the same formalism based on classes and properties. Classes represent concepts (like an Expression, a title, a key, a CD-track, etc.) or objects (items). Properties link classes to one another. This forms a graph - a well-suited representation for a future RDF implementation.

In order to grasp the DOREMUS model, one will want to consider three things:

- our records will be transformed into a graph;
- our records will be represented as a chain of events and their "products";
- our records will be modified according to the FRBR logic.

The schema in Fig. 1 is recurrent throughout the model. It describes the creation, through an event (or rather *activity*), of an Expression. This Expression is the realization of a Work.

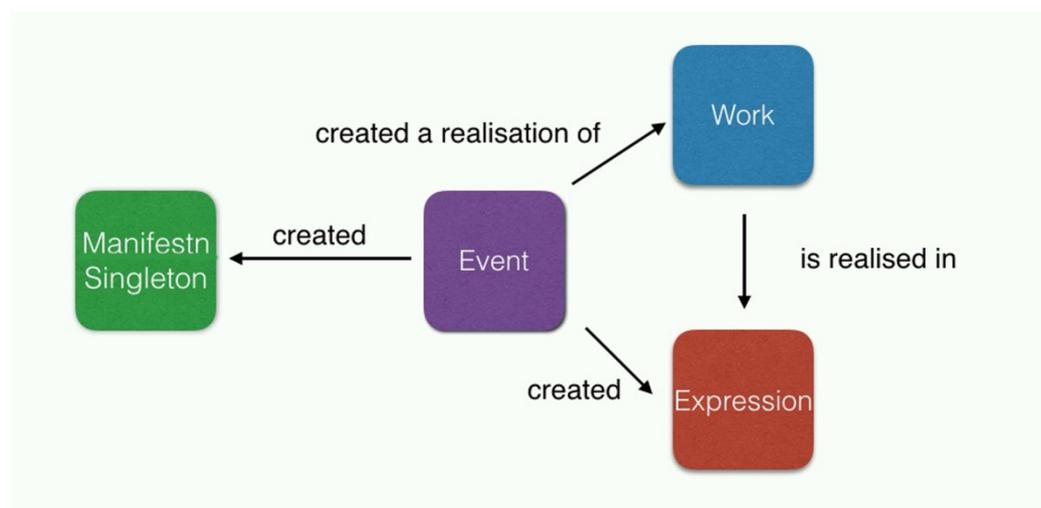

**Fig. 1:** The same event creates an Expression and a physical object

---

Essentially, anything related to the activity is described on the activity level (actor, date, place…), the intellectual result of the activity is described on the Expression level (title, genre, key…) and the physical result of the activity is described on the Manifestation Singleton level (e.g. manuscript).

Even if based on notions of Works and Expressions, as defined by FRBRer and FRBRoo, our extension of the model is nearer to the present IFLA-LRM[8] notions of Work and Expression. Indeed, all informations like title, opus number or medium of performance are mentioned at the Expression level and none at the Work level. Works only serve as hooks, linking together various Expressions.

Publications and recordings are described according to the same triangular pattern: a specific type of work – called publication or recording work – linked to an activity and an Expression. Links between musical and publication works are similar to processes in real life: as shown in Fig. 2, a musical Expression is performed, this performance is recorded and this recording activity creates a Recording which is incorporated in a Publication Expression, e.g. a CD.

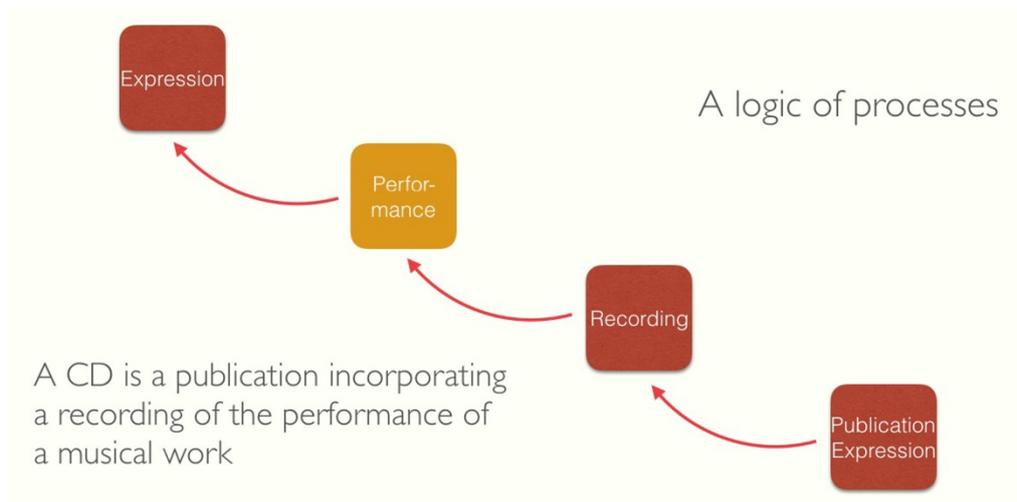

**Fig. 2:** A logic of processes, very similar to real-life processes

The concept of performance is defined by FRBRoo as a past activity which does not create but incorporates (among others) a musical Expression. This definition leaves questions unanswered: how do we deal with future events? What if the performance did not perform an Expression, but instead created one? How can we describe the result of the performance if it somehow modified the content of the original Expression – like in the genre, rhythm or instrumentation?

First, DOREMUS makes possible to describe future events thanks to new classes and properties. Then, it allows the performance to create a specific Expression (called "Performed Expression") which makes it possible to describe changes between an original musical work and its performed version. This new Expression is very useful to describe Jazz improvisations as well as non-western music which do not fit in the WEMI pattern of FRBRer (for example, the notion of musical Work can hardly be applied to Pygmy music or to an Indian Râga). Indeed, this Performed Expression may or may not have a relationship with a musical expression created by a composer.

Let us take an example of a bibliographic record describing a CD of John Coltrane's interpretation of *My Favorite Things*[9] as illustrated in Fig. 3. We start with the creation of an Expression entitled *My Favorite Things* by Rogers & Hammerstein in 1959 for the musical *The Sound of Music.* Then we have a performance by John Coltrane's Quartet (plus Eric Dolphy) on June 2, 1962, at Birdland, NYC. This performance "performed" the original song and created a new Expression, also entitled *My Favorite Things.* Its genre and duration are different, as well as its instrumentation. The performance was recorded by Vee Jay records which would be the actor of the recording activity (and probably of the first editing and the first publication). The Recording Event created an Expression, called Recording.

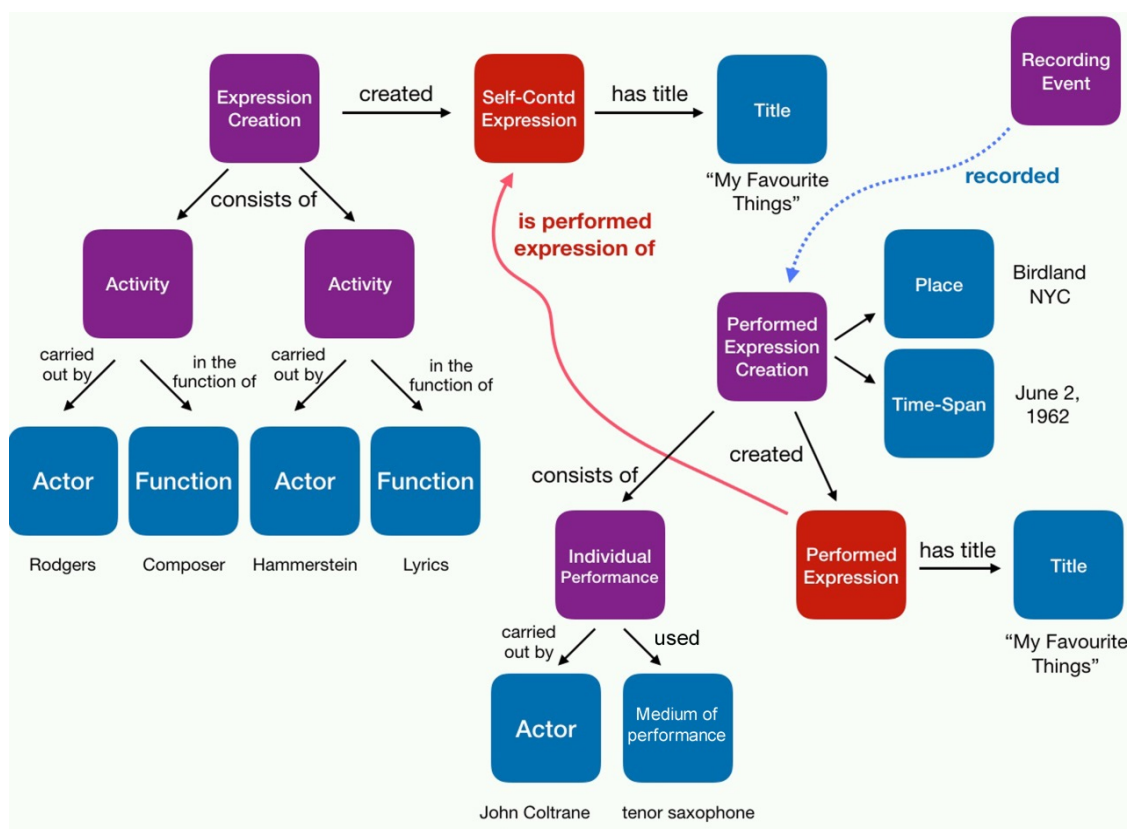

**Fig. 3:** Coltrane's performance creates a Performed Expression which *is performed expression of* the original composition by Rogers & Hammerstein. The performance is recorded by a Recording Event.

This is just an example of the kind of information that can be modeled with DOREMUS.[10] The complete documentation of the model is being finalized. A draft can be found at the following address: http://data.doremus.org/ontology/. The ontology was also edited on Protégé[11] and a working version in the RDF-Turtle format is available on Github: https://github.com/DOREMUS-ANR/doremus-ontology.

---

[9] https://www.wikiwand.com/en/My_Favorite_Things_(song).

[10] Further reading: Leresche, F.; Choffé, P.: DOREMUS: Connecting Sources, Enriching Catalogues and User Experience. IFLA 2016 (Note: the modeling of the *Performance* is an early stage version, later replaced by the *Performed Expression Creation*).

Let us see how this conceptual model was tested with the conversion of the original datasets in RDF.

## 2.3 Converting the Data to RDF

Each database differs in how to store information. In particular, there are different file formats used, representation models (meaning of the elements described), style choices (abbreviations, codes, identifiers). For this reason, the conversion of source documents to an RDF graph that follows the DOREMUS model requires different strategies to be applied. So far, three different strategies have been tested in the project. The first consists in the direct conversion of the MARC standard (in its UNIMARC and INTERMARC variants, used by the Philharmonie de Paris and the BnF respectively). Its implementation *marc2rdf* relies on a set of conversion rules between each field of the MARC file and the RDF triples to generate. For the base of concerts ("documentation sonore") of Radio France, the specificity of the representation required the development of an ad-hoc XML converter. Finally, an automated query-based SPARQL system is used to retrieve the musical works in the database.

Controlled vocabularies[12] play a key role at this stage. The conversion software includes a module that, for each text element, searches, if it exists, the best match among the concepts contained in the vocabularies, replacing literal nodes with appropriate URIs. In this way, the differences in the languages used or in the alternative forms of the labels are exceeded. As additional feature, this component is able to recognize and correct some noise that is present in the source files, like misspelled musical keys, or opus number where a catalog number is expected and vice-versa.

An additional challenge is the re-creation of the structure of the content where the information is presented in the form of plain text. Strategies based on the knowledge of librarian practices, comparison with controlled vocabularies, and text parsing, have been applied in order to extract, for example, the medium of performance from the casting notes, or the date and the publisher from the first publication note.

The output of the conversion is a dataset of music linked data available at http://data.doremus.org.

---

[11] https://protege.stanford.edu/.
[12] The entire set of DOREMUS controlled vocabularies is available under the following link https://github.com/DOREMUS-ANR/knowledge-base/tree/master/vocabularies.

## 3    DOREMUS as Linked Data

### 3.1    Vocabulary Alignment

As noted in Section 2, the universe of controlled vocabularies for music in general and particularly in the framework of DOREMUS is rich. Not only do we have multiple domains covered by distinct vocabularies, but we also often have multiple vocabularies, used by different institutions, describing one single domain. This is where vocabulary alignment comes to play. Inspired by ontology matching techniques, well-known in the Semantic Web field, we define the vocabulary alignment task as the semi-automatic identification of equivalence relations between terms of two or more thesauri (or vocabularies) covering a given field of knowledge. In DOREMUS, we focus particularly on the pairwise alignment of a set of thesauri about musical instruments (mediums of performance) and musical genres.

The alignment process unfolds in two phases: In phase 1, we produce a set of mapping candidates by the help of the automatic alignment tool LYAM++.[13] The choice of this tool, as opposed to state-of-the-art systems[14], is motivated by the fact that it is designed to handle directly SKOS vocabularies (hence not running the risk of losing information in the process of transforming SKOS to OWL), it has a rather simplified mechanism, well-suited to lightly structured resources such as our vocabularies, and it is able to handle multilingualism. In phase 2, the generated set of mappings is manually validated and enriched by experts by the help of the web environment YAM++ *on-line[15]* that we have developed particularly for this purpose.[16] The framework allows to execute an alignment by using an integrated tool and to visualize the result. Independently, it allows experts to validate the mappings of an already existing alignment (produced by the plateforme or not) and to enrich it by manually adding new mappings to the list.

### 3.2    Linking Highly Heterogeneous Music Data

Data linking, or instance matching, is defined as the task of the automated discovery of identical resources across different datasets leading to the establishment of declarative identity relations between resources. Mussorgsky's "Night on Bald Mountain" can be described in two datasets by using different identifiers and complementary information. The goal in that case would be to create a new triple that links the two Mussorgsky resources via the owl:sameAs predicate, hence bridging the two datasets. Recall that, as a result of the data transformation process, an RDF graph based on the DOREMUS model is constructed for each of our databases. Now we would like to establish links

---

between them and, to start with, we focus on linking musical Expressions described in these graphs. Many off-the-shelf data linking tools exist,[17] a recent survey can be of interest for the curious reader.[18] However, our experience showed that none of these generic tools provides satisfactory results in the world of music data. We will try to explain the major reasons for that in what follows.

First and foremost, the data that we deal with is highly heterogeneous. By heterogeneity we mean cross-datasets differences in descriptions of resources assumed to be identical. In Fig. 4 we see subsets of three large datasets: the French National Library's *data.bnf* and the two well-known knowledge graphs DBpedia and YAGO. The example attempts to illustrate the various heterogeneity types that can be observed when dealing with musical data. We categorize these heterogeneities along three main dimensions.[19]

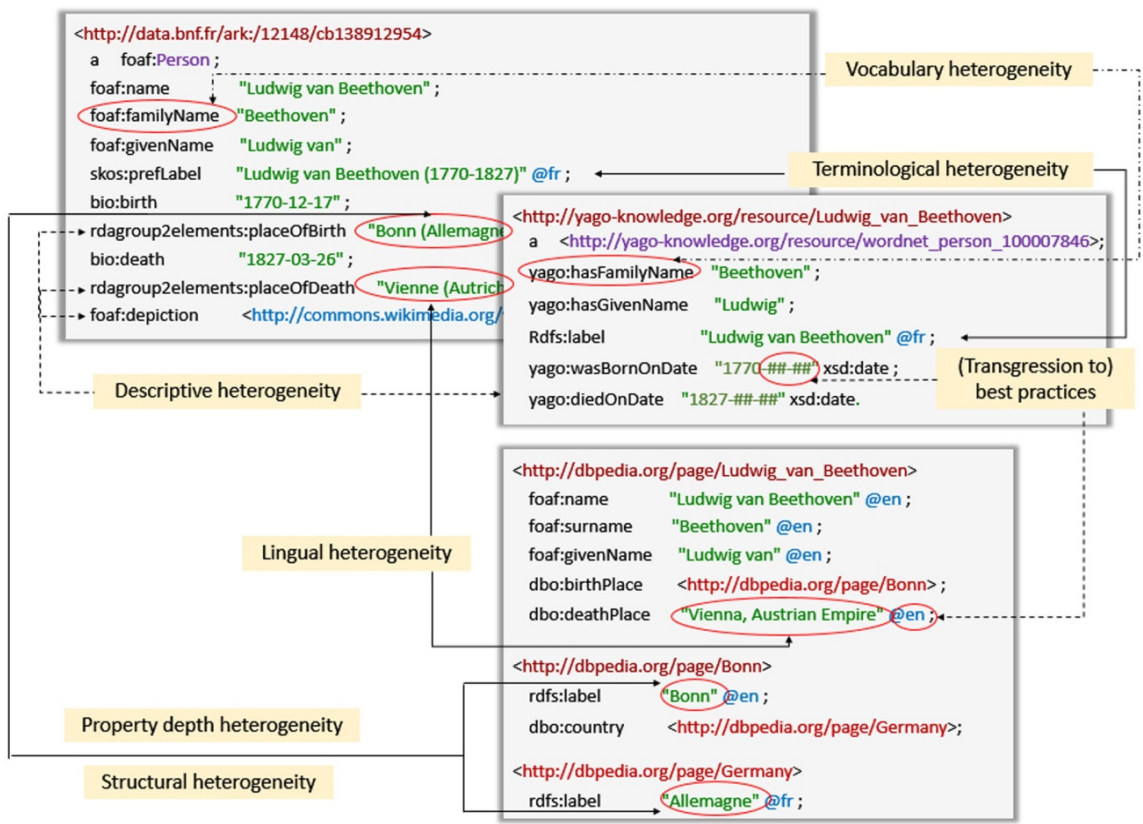

**Fig. 4:** Excerpt of descriptions of instances retrieved from **data.bnf**, **Yago** and **BDpedia**

**Value dimension:** (i) Terminological heterogeneity, expressed in minor differences in spelling or in using synonyms; (ii) Lingual heterogeneity where different natural languages are used; (iii)

---

Transgression to best practices of the semantic web; and (iv) Value type heterogeneity where the values are expressed in different formats.

**Ontological dimension:** (i) Vocabulary heterogeneity, where classes and properties are described by using different vocabularies; (ii) Structural heterogeneity: the same information is given at different levels of granularity. (iii) Property depth heterogeneity: in a given dataset, a property can be specified directly through a literal value while in another dataset the same value is accessed via a longer property chain including several triples; and (iv) Descriptive heterogeneity: lack of description in one of the two datasets.

**Logical dimension:** (i) Class heterogeneity, or the case of two resources belonging to different classes (for example, "Person" and "Composer") for which an explicit or an implicit hierarchical relationship is defined; (ii) Property heterogeneity where the equivalence between two values can be deduced after performing a reasoning task on properties (e.g., the case of inverse properties).

Most of the heterogeneities on value level are well-understood and assisted by classical methods of data unification. We focus on the ontological and logical levels that appear to be challenging for most linking systems. We also aim at automating as much as possible the process. Driven by the DOREMUS use-case (containing most of the heterogeneity types above), we developed *Legato* – a generic instance matcher, dealing with the aforementioned issues.[20]

### 3.3 The Legato Framework

The processing pipeline of Legato consists in automatically pre-processing, comparing, repairing and providing a set of identity links (a link set), as shown in Fig. 5. The system takes as an input a source and a target dataset. The following steps are performed.

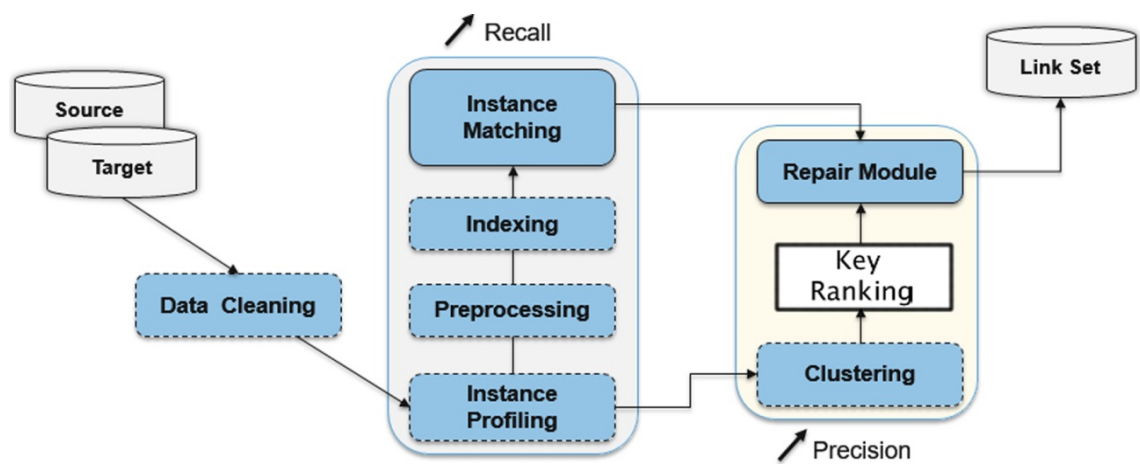

**Fig. 5:** Processing pipeline of Legato

---

[20] The Legato tool (source code, simple user interface and useful information) is available here:
https://github.com/DOREMUS-ANR/legato.

(i) Data cleaning. Certain property values make it difficult to compare the resources. Imagine the likely case of different data providers assigning different identifiers to equivalent resources across datasets (e.g., the records of a musical work in the catalogs of two libraries).

(ii) Instance profiling allos to represent each resource by a sub-graph considered relevant for the comparison task; (iii) Instance indexing and matching. These steps aim at generating a large pool of mapping candidates, guaranteeing high *recall* (the portion of all correct mappings that has been detected). Indexing techniques are applied on the resources allows to represent them as textual documents containing the values of their properties collected at a given predefined depth of the RDF graph and considered relevant for the description of a resource (Fig. 6). In that way, a work will be represented by a set of keywords coming from the RDF description of its resource. Seamlessly, this allows to compare instances in a way in which text documents are compared in a classical information retrieval framework.

(iv) Post-processing step. This step aims at reducing the false positives rate and increasing precision (the portion of all detected mappings that is correct). We apply hierarchical clustering and key identification and ranking aiming to disambiguate highly similar, though distinct pairs of works generated as candidates in steps (ii) and (iii).

The results obtained on our manually constructed benchmark data[21] showed an improved performance of Legato as compared to state of the art systems.[22]

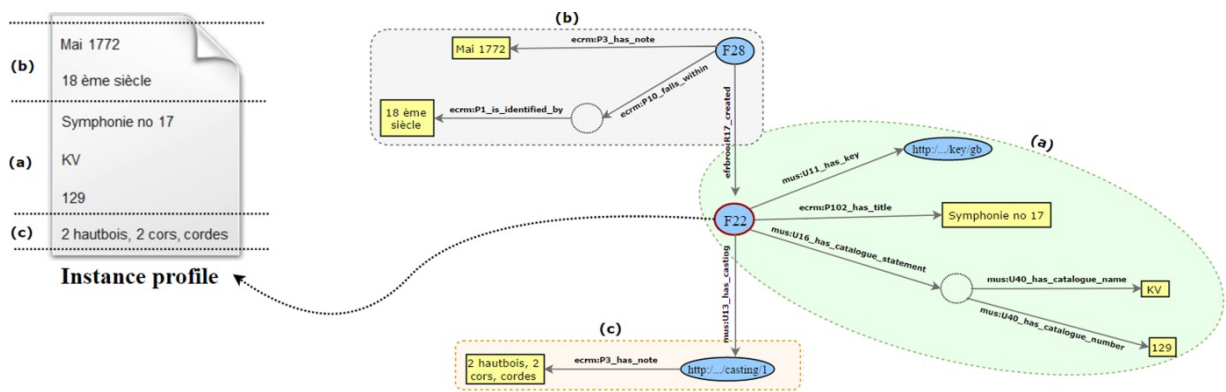

**Fig. 6:** A musical work from the DOREMUS graph represented as a text document

### 3.4 Ongoing Work and Challenges

Remember that in our context we are dealing with five large RDF graphs that need to be integrated. The solution that we opt for is creating a central pivotal graph containing the mathematical union of all music works referenced in the catalogues of the partner institutions - a sixth RDF graph equipped with owl:sameAs links to the local graphs generated by *Legato*. Further, parts of the DOREMUS

---

graphs potentially overlap with well-established datasets on the web, such as DBpedia or Musicbrainz. We will, therefore, establish links to these knowledge graphs, thus anchoring DOREMUS onto the web of data. Currently, we are working on the extraction of named entities (such as names of composers, editors, cities and orchestras) out of the values of the comment-like properties. This will help structure and enrich the information contained in the DOREMUS graphs and will contribute to a better disambiguation / comparison of resources during the data linking process.

## 4    User Needs as a Priority

Making easier for the web users to find information about music is the common goal of every task described above. This question is a fundamental principle underpinning all aspects of the project.

### 4.1    What Is Needed and Expected by the Web Users?

In order to collect as much information as possible about musical data users, and about their search and browsing practices, needs or even demands, two studies have been conducted through semi-structured interviews and usage observations. The aim was less to provide a significant typology of usages or users than to gather various kinds of needs and practices to be addressed in Doremus' achievements.

The first study took place at the very beginning of the project. About 80 people were interviewed. All of them listen digital music, and belong to at least one of these categories: (amateur or professional) musicians; music mediators and broadcasters, such as radio programmers or librarians; general public. The main topics in the interviews involve listening contexts (time, venue, support, device, sociability) and collection (set gathering, conservation, sharing, recommendation, discovery).

All the respondents spend several hours a day to listen music. The professionals tend naturally to give their exclusive attention to the music they are listening, but the line between professional and leisure listening is narrow, and all the categories of respondents take advantage of their free time to listen music. But most of them select different kinds of music, different devices or even supports, regarding their activities or the places where they listen: rhythmic music for sport, classical instrumental pieces and headphone to study, low quality MP3 recordings in public transports. Remarkably, the group listening is considered as ambient music more than as the formalized rite described by Maisonneuve.[23]

A lot of people are concerned about access, too. Most of them are users of online platforms like Spotify, Deezer, Soundcloud or mainly (and surprisingly) the video-sharing website Youtube, because "you can find everything on Youtube!" Many people play only dematerialized audio files, although they sometimes own the physical medium, or recreate it. Nevertheless, even when remote supports are favored, users value the music files retrieval on a local device, even if that means getting past the rules.[24] In addition to the pure technical interest in offline access, the files storage prevents the feeling of data or access loss (Alexandria's syndrome[25]) and seems to give an impression of gathering a collection, which is not the case in an online playlist.

---

[23]  Maisonneuve, Sophie: La constitution d'une culture et d'une écoute musicale nouvelles. Le disque et ses sociabilités comme agents de changement culturel dans les années 1920 et 1930 en Grande-Bretagne. In: Revue française de musicologie 88(1) (2002) 43–66.

To discover new pieces, even professionals and musicians do not seem to trust traditional mediators like critics. The human dimension is fundamental, but the welcomed advices come from the close social network. For the general public, the online merchants and music platforms provide additional interesting suggestions, but they remain critical of recommendation algorithms, seen as impersonal and repetitive.

Most of the interviewed persons are disappointed and even frustrated by the metadata poor quality. A precise query is unlikely to get an accurate answer, no matter the type of information that is selected: composer, interpreter, title, music style… It may sometimes foster serendipity and discovery, but remain beyond the control of the user.

The second survey involves 30 users of online music platforms. The results of this study are currently being processed. The study focused on the fields users should use to search a platform and on the information types they would follow to discover new music pieces. The achieved data should be useful to design the functional interfaces for the Overture search engine.

### 4.2   Overture: an Exploratory Search Engine

We developed the first version of *Overture* (Ontology-driVen Exploration and Recommendation of mUsical REcords), a prototype of an exploratory search engine for DOREMUS data. Overture[26] is developed as a modern web app, implemented with Node.JS and Angular. The application makes requests directly to our SPARQL endpoint and provides the information to the end-user with a nice interface. At the top of the user interface, the navigation bar allows the user to navigate between the main concepts of the DOREMUS model: expression, performance, score, recording, and artist. The challenge is in giving to the final user a complete vision on the data of each class and leaving him/her understand how they are connected to each other.

Fig. 7 represents Beethoven's *Sonata for piano and cello n.1*. Aside from the different versions of the title, the composer and a textual description, the page provides details on the information we have

---

about the work, like the musical key, the genres, the intended medium of performance, the opus number. When these values come from a controlled vocabulary, a link is present in order to search for expressions that share the same value, for example, the same genre or the same musical key. A timeline shows the most important events in the story of the work (the composition, the premiere, the first publication). Other performances and publications can be represented below. The background is a portrait of the composer that comes from DBpedia. It is retrieved thanks to the presence in the DOREMUS database of owl:sameAs links. These links come in part from the International Standard Name Identifier (ISNI) service, in part thanks to an interlinking realised by matching the artist name, birth and death date of artist of DOREMUS with the ones in DBpedia.

The page embeds also some structured markup, for helping the search engines to understand and index the content. The markup consists in a Schema.org version of DOREMUS metadata, mapped and simplified.

**Fig. 7:** How an expression in represented in Overture

The richness of the DOREMUS model offers to the end user the chance to perform a detailed advanced search, and retrieve the list of the works that exactly match the chosen filters. All expressions are searchable by facets that include the title and the composer, but also the key, the genre, the detailed casting. The use of hierarchical properties in the controlled vocabulary for genres and medium of performance, allows the smart retrieval not only of the entity that matches exactly the chosen value (i.e. "Strings, bowed"), but also any of its narrower concepts (i.e. "violin", "cello", etc.). The list of expressions is automatically updated as soon as the user modifies the parameters. In this way it is possible to search very specific information: in Fig. 8, the user is searching for all the sonatas (genre) that involves a clarinet and a piano (mediums of performance).

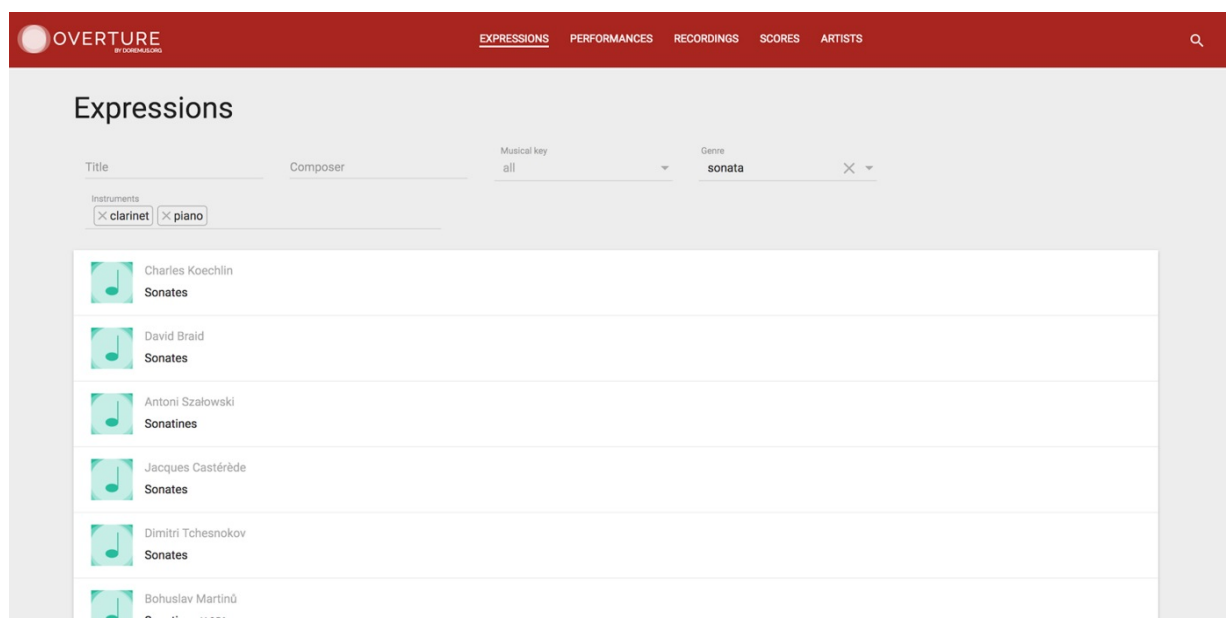

**Fig. 8:** The list of expressions in Overture filtered by genre and medium of performance

## 4.3   Towards Recommendation Systems

Our starting point for the study of a suitable algorithm for recommendation consists in two human-made collections: playlists from Radio France web radios and concert programs from the Philharmonie de Paris. While the latter can reveal very strong connection between works (they are played during the same concert, generally with the same casting), the former is characterized by a logical succession of tracks following an editorial action made by experts for an amateur public and aiming at discovering new works. These collections of works can be used as a ground truth for machine learning algorithm; the goal consists in understanding, starting from a defined work, which features (genre, casting, composer, etc.) should have the next one to listen. Here again, the richness of DOREMUS data is a good advantage. From one side, it provides a great number of features to feed the computation. From the other one, it allows to understand how the works in the recommendation path are connected each other, explaining explicitly the provided recommendation.

An experiment that involves the recommendation by features and the explanation of the recommendation, has been realised in the CityMUS,[27] a web application that gives to the user the experience of a walk in the city with the most suitable soundtrack, on the base of the urban context. The application relies on a recommender system that searches for paths in a knowledge graph between nearby places and music composers, making use of a combination of DBpedia and domain-specific datasets.

## 5    Conclusion

As a lot of countries are changing their cataloguing rules for RDA in order to facilitate the search on works and to open up the data of their libraries, we can see that the researches conducted by the DOREMUS team are closely tied to the current concerns of the library world.

The first results of the project show that it's possible to make an elaborate knowledge graph from heterogeneous and relatively flat datasets. Using a FRBR and event oriented ontology brings really new and impressive possibilities in manipulating complex datasets such as musical ones. It makes possible to create new kinds of application more accurate for the user.

DOREMUS is going to end by the spring 2018. In the meantime, the project ontology should be settled down, every original datasets converted and interlinked, several controlled vocabularies dedicated to music published (including a list of musical work authorities). The implementation and testing of the recommandation tool should also be completed.

Making our work reusable is our priority. For that reason one can find on line our documentation with some tutorials, and the code of our tools: http://data.doremus.org/.

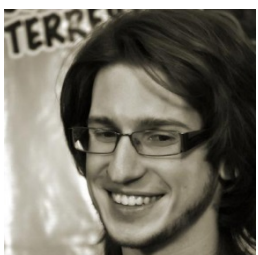

**Pasquale Lisena**
EURECOM, Sophia Antipolis, France
pasquale.lisena@eurecom.fr

---

[27] https://citymus.doremus.org.

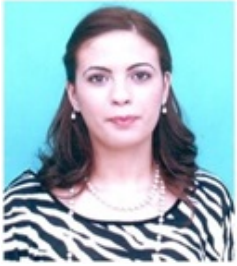

**Manel Achichi**
University of Montpellier / LIRMM, Montpellier, France.
manel.achichi@lirmm.fr

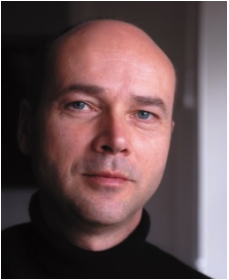

**Pierre Choffé**
Bibliothèque Nationale de France, Paris, France.
choffepierre@gmail.com

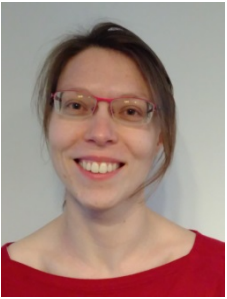

**Cécile Cecconi**
Philharmonie de Paris
Département Éducation et Ressources
221 avenue Jean Jaurès
75019 Paris
ccecconi@cite-musique.fr

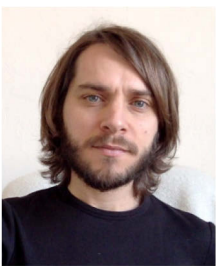

**Konstantin Todorov**
University of Montpellier / LIRMM, Montpellier, France.

konstantin.todorov@lirmm.fr

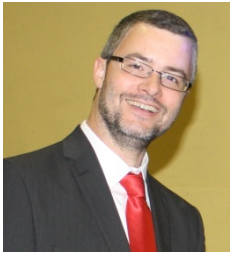

**Bernard Jacquemin**
Univ. Lille, EA 4073 - GERiiCO,

F-59000 Lille, France

bernard.jacquemin@univ-lille3.fr

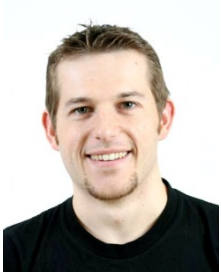

**Raphaël Troncy**
EURECOM - Campus SophiaTech

450 Route des Chappes, CS 50193

F-06904 Biot Sophia Antipolis cedex, France

raphael.troncy@eurecom.fr